\documentclass[12pt,a4paper]{article}
\usepackage{epsfig,amssymb}

\newcommand{\be}{\begin{equation}}
\newcommand{\ee}{\end{equation}}
\newcommand{\bea}{\begin{eqnarray}}
\newcommand{\eea}{\end{eqnarray}}

\newcommand{\pa}{\partial}
\newcommand{\para}{\parallel}

\newcommand{\R}{{\bf R}}
\newcommand{\Z}{{\bf Z}}

\newcommand{\s}{{\bf S}}

\newcommand{\bJ}{\bar{J}}

\newcommand{\cC}{{\cal C}}
\newcommand{\cF}{{\cal F}}
\newcommand{\cZ}{{\cal Z}}

\newcommand{\mg}{{\mathfrak g}}
\newcommand{\mh}{{\mathfrak h}}
\newcommand{\rU}{{\rm U}}

\newcommand{\rSU}{{\rm SU}}
\newcommand{\rSO}{{\rm SO}}

\newcommand{\msu}{\mathfrak{su}}

\newcommand{\dfrac}[2]{\displaystyle\frac{\mathstrut #1}{\mathstrut #2}}

\begin{document}

\thispagestyle{empty}
\addtocounter{page}{-1}

\begin{flushright}
\end{flushright}

\vskip 1cm
\begin{center}
\Large \bf A note on singular D-branes in group manifolds 
\end{center}

\vskip 1cm

\begin{center}
\large Taichi Itoh ~and~ Sang-Jin Sin
\end{center}

\begin{center} 
\it Department of Physics, Hanyang University, Seoul 133-791, Korea
\end{center}

\vskip 1cm

\begin{abstract}

After reviewing D-branes as conjugacy classes and various charge 
quantizations (modulo $k$) in WZW model, we develop the classification 
and systematic construction of all possible untwisted D-branes in Lie 
groups of A-D-E series. D-branes are classified according to their positions 
in the maximal torus. The moduli space of D-branes is naturally 
identified with a unit cell in the weight space which is exponentiated 
to be the maximal torus. However, for the D-brane classification, one may 
consider only the fundamental Weyl domain that is surrounded by the 
hyperplanes defined by Weyl reflections. We construct all the D-branes 
by the method of iterative deletion in the Dynkin diagram. 
The dimension of a D-brane always becomes an even number and 
it reduces as we go from a generic point of the fundamental domain to its 
higher co-dimensional boundaries. Quantum mechanical stability requires 
that only D-branes at discrete positions are allowed. 
 
\end{abstract}

\vspace*{\fill}
\hrule

\vskip 0.2cm
\noindent
taichi@hepth.hanyang.ac.kr,\\
sjs@hepth.hanyang.ac.kr

\baselineskip=18pt
\newpage
\renewcommand{\theequation}{\arabic{section}\mbox{.}\arabic{equation}} 

\setcounter{equation}{0}
\section{Introduction}

D-branes are central elements of modern string theory.
Therefore it is important to ask what are possible D-brane configurations 
in curved space. Group manifolds provide us solvable string theory backgrounds 
in terms of current algebra and D-branes therein can be described by 
the gluing conditions \cite{AS,KO,S}, which specify how the left and right 
currents are matched along the D-branes. In WZW model, there is an intrinsic 
$B$ field in the bulk of the group manifold, as well as a U(1) gauge field 
on the D-brane such that $B+F$ is the gauge invariant quantity.
It is interesting to observe that it is the charge of $F$ field 
that is quantized. This was first observed in \cite{BDS} and explained 
in physical terms in \cite{T}. 
Later this phenomena got a purely geometric explanation \cite{S2,FS}. 
The D-brane charges and their relations to twisted K-theory were 
recognized in \cite{FrSc,MMS}. 

Though the group manifold of lower dimension can be 
applied to string theory, higher dimensional groups can also 
play a role through the coset constructions. Therefore the study of 
D-branes in general group manifolds is interesting as well. 
There is an extensive literature on this subject, however it is 
mostly on either the generic brane or the D0-brane. 
The former is of the highest possible dimension in a given group manifold, 
while the latter is of the lowest dimension. However, it is quite clear that 
there is a variety of D-branes between these two extremes as can be seen 
from recognizing D-branes as the conjugacy classes. 
In a recent paper, Stanciu \cite{S3} studied such singular D-branes 
in SU(3) case. In this paper we consider more general group manifolds 
whose Lie algebras are simply-laced (A-D-E series). 

D-branes will be classified according to their positions in the maximal torus. 
The moduli space of D-branes is naturally identified with a unit cell 
in the weight space which is exponentiated to be the maximal torus. 
However, for the D-brane classification, one may consider only the 
fundamental Weyl domain that is surrounded by the hyperplanes defined 
by Weyl reflections. 
We construct all the D-branes by the method of iterative deletion in 
the Dynkin diagram. The dimension of a D-brane always becomes an 
even number and it reduces as we go from a generic point of the fundamental 
domain to its higher co-dimensional boundaries. Quantum mechanical stability 
requires that only D-branes at discrete positions are allowed. We consider 
only the untwisted D-branes leaving the twisted case to later study.

The rest of the paper consists as follows: 
In section 2, we review the quantization of the level $k$ of WZW model and 
that of `D0-charge' modulo $k$, by considering the single-valuedness of 
the path integral of WZW model of open strings.
In section 3, we review the identification of the D-branes as 
the conjugacy classes.
In section 4, we describe the classification of the singular D-branes, 
which is the main contents of this paper. 
In section 5, we discuss the discretization of D-brane positions due to 
quantum mechanical consistency. 
In section 6, we summarize and conclude with future projects.

\setcounter{equation}{0}
\section{D-branes in group manifolds: a review} 

In this section we give a self-contained review of the quantization 
of the level $k$ and that of D$(p-2)$-charge in a D$p$-brane modulo $k$. 
The former is associated with the existence of $H$-monopoles while the latter 
with that of $F$-monopoles. We also give a brief summary of the reasoning 
why the D-branes in group manifolds are given by the conjugacy classes. 

\subsection{WZW model and $H$-monopoles} 

The sigma model action for closed strings propagating in the group 
manifold $G$ is given by Wess-Zumino-Witten (WZW) action without 
boundary terms \cite{W}:
\be
S_{closed}=\frac{k}{4\pi} \int_\Sigma 
{\rm Tr}\!\left(\pa_+ g \,\pa_- g^{-1}\right)
+\int_M H, \label{WZW}
\ee
where $k$ is a number to be identified with the level of the 
relevant affine Lie algebra and 
\be
H=\frac{k}{12\pi}\,{\rm Tr}\!\left[(g^{-1}dg)^3\right] \label{H}
\ee
is the canonical 3-form defined on a 3-dimensional extension of the 
string world sheet in target manifold, i.e, it is defined on $M\subset G$ 
such that $\pa M =g(\Sigma)$. 
For a compact semi-simple Lie group $G$, one can extend $g(\Sigma)$ to $M$ 
since there is no topological obstacle, namely, $\pi_2(G)=\{0\}$.
Normalization of the 3-form $H$ is fixed by the conformal symmetry \cite{W}. 
The symmetry of WZW model is the affine Lie algebra generated by 
chiral currents of $G_L \times G_R$ symmetry.

Since $H$ is a closed 3-form, it is locally given by $H=dB$ on $M$
allowing us to rewrite the Wess-Zumino (WZ) action as
\be
S^{WZ}\equiv \int_M H = \int_{g(\Sigma)} B =\int_\Sigma \,g^* B.
\ee
So it describes the coupling of the closed string world sheet with 
the NS $B$ field. 
Notice that the WZ action depends on how $g(\Sigma)$ is extended, 
namely it depends on our choice of $M$. In order to make the theory 
well-defined in quantum level, such an ambiguity in WZ action should be 
modulo $2\pi$. 
This requirement leads to restricting the level $k$ to be an integer; 
Imagine 3-dimensional manifolds $M_1$ and $M_2$ which share the same 
boundary $g(\Sigma)$ and are oriented to $g(\Sigma)$ in the same way, 
for instance, $M_1$ as bulk inside $g(\Sigma)$, whereas $M_2$ as bulk 
outside. The difference between the WZ actions defined on $M_1$ and $M_2$ is 
\be
\int_Z H
\ee
where $Z \equiv M_1-M_2$ is a 3-cycle in $G$. 
If the above integral is nonzero then $H$ must be a nonzero element of
$H^3(G;\R)=\R$. In fact, $\pi_3(G)=\Z$ for any compact semi-simple Lie 
group $G$ ensures that the integral is an integer multiplied by $2\pi k$, 
that is
\be
\int_Z H =2\pi k n,\quad n\in\Z,\label{pontr}
\ee
where $n$ counts the winding number of the extended map $g: Z \mapsto G$ 
which is wrapped by the 3-cycle $Z$. 
The winding number $n$ is also identified with the number of $B$ field 
monopoles (hereafter simply referred to as $H$-monopoles) inside 
4-dimensional bulk enclosed by $Z$.

The well-definedness of WZ action therefore requires that the integral 
(\ref{pontr}) takes values only on $2\pi\Z$. 
In other words, $H$ must be an element of the integer cohomology: 
\be
[H]/2\pi \in H^3(G;\Z)\label{req}.
\ee
Moreover, 
\be
k n \in \Z,\quad\forall\, n\in\Z \quad\Longrightarrow \quad k\in\Z. 
\label{levelk}
\ee
Thus the WZW model with an integer $k$ 
becomes a well-defined quantum theory. It is referred to the level $k$ 
WZW model \cite{W}. Notice that the period of $H$ actually takes 
values on $k\Z$ and not on the whole of $\Z$:
\be
[H]/2\pi \in H^3(G;k\Z). \label{modk}
\ee
In order words, $[H]/2\pi$ as an element of $H^3(G;\Z_k)$ is 0, 
where $\Z_k\equiv\Z/k\Z=\{0,1,\dots,k-1\}$.
Later, we will see that the same modularity $\Z_k$ arises 
also in the WZW model for open strings.

\subsection{WZW model for open strings and D$(p-2)$-charge} 

D-branes are hypersurfaces obtained by imposing Dirichlet boundary 
conditions to open string end points. Since each end point carries 
a Chan-Paton degree of freedom which couples to a U(1) gauge field $A$, 
the gauge invariant 2-form on a D-brane is $\cF \equiv B+F$, 
the combination of the NS $B$ field and the U(1) gauge flux $F=dA$. 

\begin{figure}
\begin{center}
\begin{minipage}[t]{3.5cm}
\centerline{\hbox{\psfig{file=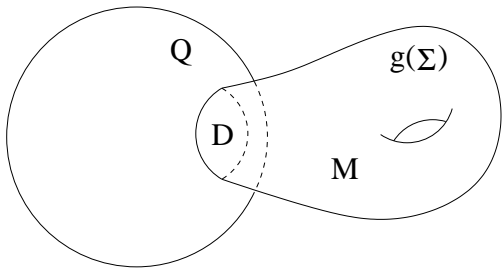,height=3.5cm}}}
\end{minipage}
\caption{An open string world sheet in a group manifold $G$. 
$g(\Sigma)$ is the open string world sheet embedded by $g(x)$ into $G$, 
while $M$ is a 3-dimensional manifold with boundary $g(\Sigma)+D$.
The 2-disk $D$ is embedded in a D-brane world volume $Q$.}
\label{fig:open}
\end{center}
\end{figure}

We can think of an open string world sheet a closed surface
by attaching some 2-disks. Since the open string end points are attached 
on D-branes, one can choose the 2-disks as to be embedded in a D-brane 
world volume $Q$. In figure \ref{fig:open}, $g(\Sigma)$ is the image of 
an open string world sheet $\Sigma$ under $g: \Sigma \mapsto G$, while $D$ is 
a 2-disk embedded in $Q$. Let $M$ be the 3-dimensional manifold whose 
boundary is $\pa M =g(\Sigma) + D$. In general, $D$ may be replaced by a set 
of disjoint 2-disks $\bigcup_i D_i$. Now we can extend the mapping $g$ 
to three dimensions to define the WZW action.

The WZW model for open strings can be described by the sigma model action 
\cite{KS}
\be
S_{open} =\frac{k}{4\pi}\int_\Sigma 
{\rm Tr}\!\left(\pa_+ g \,\pa_- g^{-1}\right)
+\int_M H-\int_D \cF, \label{BWZW}
\ee 
where the first two terms are familiar from WZW action for closed strings, 
while $\cF$ is a 2-form on $D\subset Q$ defined by $\cF=B+F$, 
with NS $B$ field and the U(1) gauge field $F=dA$ defined on the D-brane 
world volume. Apparently, the WZ action, namely the last two terms, 
depends on the choice of the 2-disks to form a closed surface out of the open 
string world sheet. If $H=dB$ globally, however, one can easily show 
this is not the case;
\be
S^{WZ} \equiv \int_M H -\int_D \cF =\int_{g(\Sigma)} B -\int_D F
=\int_\Sigma \,g^* B+\int_{\pa\Sigma} \,g^* A,
\ee
where we have used $g(\pa \Sigma)=-\pa D$ in the last equality.
If $H \ne dB$ globally, or equivalently if $\int_{Z_3} H \ne 0$ for any 
3-cycle $Z_3$, the above action for open strings has some ambiguity associated 
with the choice of $(M,D)$. 

\begin{figure}
\begin{center}
\begin{minipage}[t]{5.5cm}
\centerline{\hbox{\psfig{file=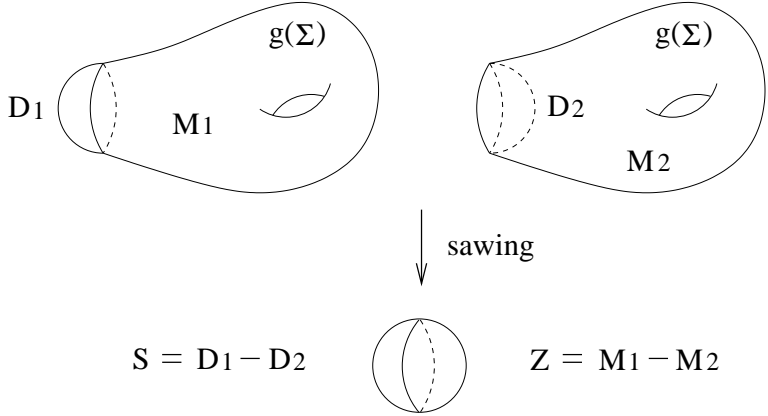,height=5.5cm}}}
\end{minipage}
\caption{Sawing $M_1$ and $M_2$ along the open string world sheet 
$g(\Sigma)$. $(M_1,D_1)$ and $(M_2,D_2)$ have the same orientation 
relative to $g(\Sigma)$.}
\label{fig:sawing}
\end{center}
\end{figure}

Suppose we have $(M_1,D_1)$ and $(M_2,D_2)$ which share the same relative 
boundary $g(\Sigma)$ modulo $Q$, say explicitly 
$\pa M_1-D_1=\pa M_2-D_2=g(\Sigma)$.
The difference between the two $S^{WZ}$'s becomes
\be
\int_Z H -\int_S \cF \equiv C \label{vari}
\ee
where $(Z,S)=(M_1-M_2,D_1-D_2)$. See figure \ref{fig:sawing}. 

\begin{figure}
\begin{center}
\begin{minipage}[t]{6.5cm}
\centerline{\hbox{\psfig{file=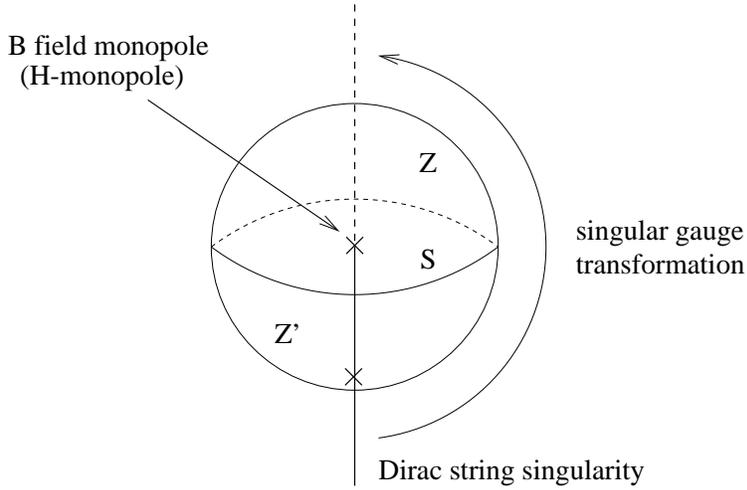,height=6.5cm}}}
\end{minipage}
\caption{$B$ field monopoles ($H$-monopoles) and Dirac string 
singularities on a 3-cycle $Z^{'}-Z$. $Z$ and $Z^{'}$ are 3-balls sharing 
the same boundary 2-sphere $S$. In the figure, only $Z^{'}$ intersects 
the Dirac strings associated with $H$-monopoles. The bulk $B$ field locally 
defined on $Z$ is different from the one defined on $Z^{'}$ by a singular 
gauge transformation.}
\label{fig:sing}
\end{center}
\end{figure}

In contrast to the closed string case, $Z$ is not a 3-cycle but a 3-ball 
with boundary $\pa Z \equiv S \subset Q$. As long as we restrict ourselves 
on the local area of $Z$, we can set $(H,\cF)=(dB,B+F)$ on $(Z,S)$ to yield
\be
C =-\int_S F, \label{D0}
\ee
which is the U(1) gauge flux through $S$ \cite{T}. 
The well-definedness of quantum theory therefore requires $C$ to take values 
on $2\pi\Z$ \cite{AS,T,FS,KS,G}. $C$ can be considered as the total flux 
of monopoles of the U(1) gauge field $A$ ($F$-monopoles) enclosed by 
the 2-cycle $S$. Hence the requirement can be interpreted as the Dirac 
quantization condition of the $F$-monopole charge. 
If we naively extend this to all over the group manifold, 
we would have a general condition
\be
[F]/2\pi \in H^2(Q;\Z). \label{Fmono}
\ee
This is true, however, only when there is no $H$-monopole. 
In fact, $\pi_3(G)=\Z$ tells us 
that there cannot exist such a $B$ field globally defined on $G$. 
To study this, let us consider 3-balls $Z$ and 
$Z^{'}$ sharing the common boundary $S$ so that $Z^{'}-Z$ form a 3-cycle 
in $G$ as shown in figure \ref{fig:sing}. 
Now suppose there exist $H$-monopoles enclosed by the 3-cycle $Z^{'}-Z$, 
i.e, $\int_{Z'-Z} H \ne 0$. 
Then, even though we can choose $Z$ as a coordinate patch 
without any singularities, 
$Z^{'}$ necessarily contains singularities due to the Dirac strings of 
$H$-monopoles. For $C^{'}$ defined on $(Z^{'},S)$, namely, 
$C^{'} \equiv \int_{Z^{'}} H -\int_S \cF$, the $B$ field in $Z^{'}$ must be 
different from that in $Z$ and both are related by a singular gauge 
transformation on their common boundary $S$. For the level $k$ 3-form $H$, 
the shift in $C$ due to the singular gauge transformation on $S$ must be 
an integer multiple of $2\pi k$ just like 
the closed string case (See Eq.\ (\ref{modk})):
\be
C^{'}-C =\int_S (B'-B)= \int_{Z^{'}-Z} H =2\pi k n.
\ee
 
Recall that $C$ is the $F$-monopole flux of the D-brane $Q$ and 
$C^{'}-C$ is just the effect of Dirac strings of the $H$-monopoles, 
therefore it should not be observed. 
This requires that both $C^{'}$ and $C$ must be equal modulo $2\pi k$:
\be
C^{'} \equiv C \quad\mbox{mod } \; 2\pi k. \label{modk2}
\ee
Therefore the $F$-monopole charge (the value of $C$) is defined up to 
modulo $k$: 
\be
[F]/2\pi \in H^2(Q;\Z_k). \label{Fmono2}
\ee
This is the desired result \cite{AS,FS}.
In D$p$-branes, the $F$-monopole charge can be identified as 
D$(p-2)$-charge, since the WZ term of Dirac-Born-Infeld (DBI) action 
gives the relevant term 
\be
\int_{p+1}C_{p-1}\wedge F, 
\ee where $C_{p-1}$ is the RR gauge potential 
that couples to D$(p-2)$-branes. 
In D2-branes, it can be identified as D0-charge. 
 
To summarize, the level $k$ boundary WZW model becomes a well-defined 
quantum theory only if the D$(p-2)$-charge of a D$p$-brane is 
modulo $k$ integer valued. 

\setcounter{equation}{0} 
\section{D-branes as conjugacy classes} 

In this section we will see how the boundary 
conditions in WZW model for open strings provide D-branes as conjugacy 
classes. The chiral $G_L \times G_R$ symmetry of WZW model is generated by
the left and right chiral currents 
\be
J = -\pa_+ g g^{-1},\quad \bJ = g^{-1} \pa_- g,
\ee
at each point $g\in G$ in target space. The currents induce translations 
on the group manifold $G$ 
induced by left and right multiplications;
\be
g \mapsto l g\hspace{0.5pt} r^{-1}
\quad\mbox{with}\quad(l,r)\in (G_L,G_R) 
\ee
where $(l,r)$ are given by exponentiating $(J,\bJ)$ at each point 
$g\in G$. 

In the boundary WZW model, we are interested in the string world sheet 
boundary conditions which preserve half of the chiral $G_L \times G_R$ 
symmetry \cite{KO}. 
Such a boundary condition may be given by equating $J$ with $\bJ$ up to 
automorphisms $\Omega$ of the Lie algebra $\mathfrak g$, namely the 
gluing condition \cite{AS,FFFS,S}:
\be
J =\Omega(\bJ)\quad \mbox{at}\quad\sigma=0.\label{glue}
\ee
This condition restricts the isomorphic mapping $f$ to $f_\omega\!:G\to G$ 
induced by
\be
g \mapsto \omega(r) g\hspace{0.5pt} r^{-1} \quad\mbox{with}\quad r \in G_R,
\ee
where $\omega$ acting on $r \in G_R$ is generated by the automorphism 
$\Omega$ acting on $\bJ$. Since the boundary mapping $f_\omega$ translates 
open string end points, the string world sheet boundary sweeps out the 
invariant submanifold of $G$ under $f_\omega$, that arises as a twisted 
conjugacy class:
\be
\cC_\omega (h)=\{\omega(g)hg^{-1}~\mbox{with}~g \in G\},
\label{conj}
\ee
which can be regarded as a D-brane fixed at a point $h\in G$.
It was shown in \cite{AS} that the gluing condition 
(\ref{glue}) actually provides boundary conditions (BC's) 
so as to identify a D-brane in a group manifold as a conjugacy class 
(\ref{conj}) and it was generalized to twisted cases in \cite{FFFS,S,S2}.
In this paper our main concern is the untwisted case so we set $\Omega=1$, 
$\omega=1$.

One can derive the Dirichlet and Neumann BC's from the 
gluing condition (\ref{glue}) as follows.
The open string gluing condition $J=\bJ$ can be rewritten as 
\be
(1+Ad_g)\pa_\tau g =(1-Ad_g)\pa_\sigma g,
\ee 
where $Ad_g(v)=gvg^{-1}$ is the adjoint action of the Lie group on 
the tangent vector $v$. Now we decompose the tangent space at $g\in G$ into 
parallel and orthogonal directions to the adjoint action:
\be
T_g G =T_g^\perp G \oplus T_g^\para G.
\ee
Then $Ad_g=1$ in $T_g^\perp G$ by definition and we get 
\be
(\pa_\tau g)^\perp = 0. \label{gl2}
\ee 
Introducing the local coordinates $X^a$ of $G$ near $g$ and the corresponding 
basis of the tangent space $e_a:=\pa_{X^a}g$, any tangent vector can be 
written as
\be
\delta g= e_a \delta X^a. 
\ee
Inserting this in Eq.\ (\ref{gl2}), we get the Dirichlet boundary condition 
$\pa_\tau X^\perp=0$. 
For the parallel directions we get the `Neumann' BC given by 
\be (\pa_\sigma g)^\para = \frac{(1+Ad_g)}{(1-Ad_g)}(\pa_\tau g)^\para.
\ee
Notice that the operator $(1-Ad_g)^{-1}(1+Ad_g)$ is defined only on parallel 
directions. Using the coordinates $X^a$ of the group manifold
and the tangent space basis $e_a$'s, these BC's can be rewritten as 
\bea
\mbox{Dirichlet BC's}:
&& \left.\pa_\tau X^a_\perp \right|_{\sigma=0} =0, \label{DBC}\\
\mbox{Neumann BC's}:
&& \left.\pa_\sigma X^a_\para -\cF^a_{~b}\,\pa_\tau 
X^b_\para \right|_{\sigma=0}=0 , \label{NBC}
\eea
where $\cF^b_{~a}$ is defined by 
\be
[(1-Ad_g)^{-1}(1+Ad_g)](e_a)=\cF^b_{~a} e_b .
\ee 
This establishes the fact that D-branes are given by the conjugacy 
classes \cite{AS}. The Neumann BC's in Eq.\ (\ref{NBC}) tell us that 
the gauge invariant 2-form $\cF$ on D-branes is given by \cite{AS} 
\be
\cF=\frac{k}{8\pi}\,{\rm Tr}\left[(dgg^{-1})^\para 
\,\frac{1+Ad_g}{1-Ad_g}\,(dgg^{-1})^\para\right].
\label{giF}
\ee
Almost identical argument can be applied to the twisted case; Twisted 
boundary condition gives the D-branes as the twisted 
conjugacy classes \cite{FFFS}. 

It can be shown that any conjugacy class contains a point 
in the maximal torus, that is the abelian subgroup generated by the Cartan 
subalgebra $\mathfrak{h}\subset\mathfrak{g}$. So the conjugacy classes can be 
parameterized by the points in the maximal torus. 
We use $\cC(h)$ to denote the conjugacy class that passes through the point 
$h$ on the maximal torus. The D-branes are thereby classified 
according to their positions in the maximal torus. 

Let $\cZ(h)$ be the symmetry group of $h$, $\cZ(h)=\{ g\in G|\,ghg^{-1}=h\}$. 
It is also called the centralizer of $h$.
Then the D-brane or the conjugacy class passing $h$ is given by 
the homogeneous space 
\be
\cC(h)= G/\cZ(h).
\ee
The dimension of a D-brane thus depends on the symmetry group of $h$. 

If $h$ is a generic point in $T $, 
$\cZ(h)$ is obviously given by $T $ itself and its 
conjugacy class $\cC(h)$ arises as a quotient space $G/T $ 
and this is the D-brane of maximal dimension \cite{FFFS}. 
For $G=\rSU(2)$, the conjugacy class $\rSU(2)/\rU(1)$ is a spherical D2-brane. 
For $G=\rSU(3)$ the generic D-brane is a D6-brane given by 
$\rSU(3)/\rU(1)^2$. 

If $h$ is a singular point of $T $ where some of its U(1) 
subgroups are enhanced to SU(2), the centralizer $\cZ(h)$ becomes larger 
than $T $ including the SU(2)'s resulting in a D-brane with lower dimension.
One can think of such a D-brane as a singular limit of the maximal D-brane. 
For instance, for $G=\rSU(2)$, the point-like D-branes exist on the north 
and the south poles of $\s^3$ where the spherical D-brane shrinks to a point.
To classify the D-branes, 
we need to look at the enhancement of centralizer groups $\cZ(h)$ for 
all $h\in T$. That is the topic of the next section.

\setcounter{equation}{0}
\section{Singular D-branes and their classification}


In this section we will develop the general structure of the centralizer 
enhancement and corresponding D-branes as a consequence. 
In order to avoid the complication coming from the Dynkin 
diagram symmetry associated with outer automorphisms, we will restrict 
ourselves to regular conjugacy classes without twist.
The argument in $G=\rSU(3)$ can be found in \cite{S3}. 
As we discussed before the conjugacy class is isomorphic 
 to the group manifold modulo the centralizer. 
 Therefore we need to look at the centralizer of an
 arbitrary point $h$ in the maximal torus, which can be given by 
exponentiating an element $X$ in the Cartan subalgebra $\mh\subset\mg$. 
More explicitly, $h\equiv e^{2\pi iX}$ and $X$ is parameterized 
by a weight vector $\vec{\psi}$ such that
\be
X=\vec{\psi}\cdot\vec{H}\quad\mbox{where}\quad 
\vec{H}=(H_1,\dots,H_r) \quad\mbox{with}\quad H_i \in \mh,
\label{x1}
\ee
where $r \equiv {\rm rank}\, \mg$. 

A given Lie algebra $\mg$ of dimension $d$ and rank $r$ has the Cartan 
decomposition $\mg =\mh\oplus(\bigoplus_\alpha \mg_\alpha)$;
\bea
[\vec{H}, \vec{H}\,] &=& 0,\nonumber\\ 
\left.[\vec{H}, E_{\pm \alpha_i}]\right. 
&=& \pm\vec{\alpha}_i E_{\pm \alpha_i},\nonumber\\ 
\left.[E_{\alpha_i}, E_{-\alpha_j}]\right. &=& \delta_{ij}\,
\vec{\alpha}_i \cdot \vec{H},
\label{Lie}
\eea
where $\alpha_i$'s ($i=1,\dots,\frac{1}{2}(d-r)$) denote positive
roots and corresponding root vectors in weight space are given by
$\vec{\alpha}_i$'s. The first $r$ roots $\{\alpha_1,\dots,\alpha_r\}$
denote simple roots. Note that the second equation in Eq.\ (\ref{Lie})
fixes the normalization of the Cartan generators $\vec{H}$ so that
they scale as $\vec{\alpha_i}$. 
Now we introduce the scale invariant generators of
 Lie algebra $\mathfrak{su}(2)_{\alpha_i}$:
\be
H_{\alpha_i}\equiv\frac{\vec{\alpha}_i \cdot \vec{H}}
{|\hspace{1pt}\vec{\alpha}_i|^2},\quad
e_{\pm \alpha_i}\equiv\frac{E_{\pm 
\alpha_i}}{|\hspace{1pt}\vec{\alpha}_i|}.
\label{norm}
\ee
These can be identified respectively with the spin operators of $\rSU(2)$: 
\be
J^3,\quad J^\pm \equiv\frac{1}{\sqrt{2}}(J_1 \pm iJ_2).
\ee

One first notice that 
$\exp\,(2\pi i n J_3)$ with arbitrary integers $n$ commute with 
all generators of the Lie algebra $\msu(2)$ and so is 
$\exp\,(2\pi i nH_{\alpha_i})$ in the subgroup $\rSU(2)_{\alpha_i}$.
This can be checked by explicit calculation using the basis given 
in appendix B. 
Let's recall that $h = \exp(2\pi iX) \in T$ and $X=\vec{\psi}\cdot \vec{H}$. 
Generic points on the maximal torus $T$ have the same centralizer 
$\rU(1)^r$, where each $\rU(1)$ is generated by $H_{\alpha_i}$.
There are singular points in $T$ 
where some of $\rU(1)$'s are enhanced to $\rSU(2)$'s. These points 
are the image of a hyperplane in Cartan subalgebra under the 
exponentiation. 
The intersection of $k$ such hyperplanes 
corresponds to higher singular points where $k$ U(1)'s 
in the centralizer are enhanced to SU(2)'s.
We now describe these enhancements more concretely. 
 
We first decompose $X\in \mh$ into $H_{\alpha_i}$ direction and 
its orthogonal complement: 
\be
X\equiv (\vec{\alpha}_i\cdot\vec{\psi}) H_{\alpha_i}+X^\perp,
\label{proj}
\ee
Then 
\be
[X^\perp, E_{\pm\alpha_i}]=0,\ee 
as can be easily checked by using the Lie algebra in Eq.\ (\ref{Lie}) and 
the fact that 
$\vec{\psi}-\vec{\alpha}\,(\vec{\psi}\cdot\vec{\alpha})/|\vec{\alpha}|^2$ 
is orthogonal to $\vec{\alpha}$ for any $\vec{\psi}$. 
The exponentiation of $X$ is therefore factorized such that
\be
h\equiv e^{2\pi i X}=h^\perp 
\exp\left[2\pi i(\vec{\alpha}_i\cdot\vec{\psi}) H_{\alpha_i}\right],
\ee
where $h^\perp \equiv e^{2\pi i X^\perp}$ commutes with $\msu(2)_{\alpha_i}$. 
Then $h$ commutes with $E_{\pm \alpha_i}$ when $\vec{\psi}$ 
is located on any of the hyperplanes defined by 
\be
\vec{\alpha}_i\cdot\vec{\psi}=n, \quad n \in\Z.\label{4.8}
\ee 
Such hyperplanes are perpendicular to the root vector $\vec{\alpha}_i$ 
since they are all parallel to the hyperplane defined by 
$\vec{\alpha}_i\cdot\vec{\psi}=0$.
On those hyperplanes, $\rU(1)_{\alpha_i}$ 
in the centralizer is enhanced to $\rSU(2)_{\alpha_i}$. 
Consequently $\cZ(h)$ becomes $\rSU(2)\otimes\rU(1)^{r-1}$. Notice that
the rank of the centralizer is preserved under these enhancements.

Now we introduce the fundamental weight vectors
$\{\vec{\mu}_1,\dots,\vec{\mu}_r\}$ as a basis of weight space. 
We use following normalization \cite{Geo}: 
\be
\frac{2\,\vec{\alpha}_i \cdot\vec{\mu}_j}
{|\hspace{1pt}\vec{\alpha}_i|^2}=\delta_{ij}, \label{fund}
\ee
where $\vec{\alpha}_i$'s are restricted to simple roots only. 
Setting the universal length of all the root vectors $\sqrt{2}$, 
each root vector can be identified with a weight vector generating 
the adjoint representation of the group $G$. 
Under the decomposition
\be
\vec{\psi}=\sum_{i=1}^r \psi_i \,\vec{\mu}_i,\label{Hm}
\ee
the coordinates $\psi_i$'s can be calculated to be
\be
\psi_i =\vec{\alpha_i}\cdot\vec{\psi}. \label{psi}
\ee
The hyperplane $\vec{\alpha}_i\cdot\vec{\psi}=n \in\Z$ of the enhanced 
symmetry $\rSU(2)_{\alpha_i}$ is specified by $\vec{\psi}$'s whose $\psi_i=n$. 
If a positive root $\beta$ has the simple root decomposition 
$\beta=\sum_i p_i \alpha_i$, then the hyperplanes of $\rSU(2)_\beta$ 
enhancement are given by 
\be 
\sum_i^r {p_i}\psi_i =n,\label{betap}
\ee
for some integer $n$. Note that the coordinates (\ref{psi}) defined for the 
simple roots can be extensively used for non-simple roots also. 
For example one can define a coordinate 
$\psi_\beta \equiv \vec{\beta}\cdot\vec{\psi}$ and ensures that $\psi_\beta=n$ 
yields the same hyperplane as in Eq.\ (\ref{betap}). 

Denoting the hyperplanes with $\rSU(2)_{\alpha_i}$ symmetry by 
\be
P_{\alpha_i,\,n_i}= \{\vec{\psi} |\,\psi_i = n_i,\,n_i\in\Z\},
\ee 
the hyperplanes for all the positive roots divide the whole of weight space 
into the fundamental domains surrounded by the hyperplanes. 
The integral lattice is defined by the inverse image of the identity 
of $G$ under the exponentiation \cite{adams}. 
Define the unit cell of a lattice by dividing the weight space by the lattice. 
Then the moduli space, or the maximal torus, is in one to one 
correspondence with the unit cell of the integral lattice. 
However, the problem of the symmetry enhancement is not directly related 
to the periodicity of 
the integral lattice but other concept, so called the central lattice. 
It can be defined as  the intersection points of the $r$ hyperplanes 
corresponding to the simple roots.
This lattice is mapped to the center of the group by the exponential map, 
which justifies the name. 
For simply-laced cases, Eqs.\ (\ref{4.8}), (\ref{fund}) show that this 
lattice is generated by the fundamental weight vectors. 
Therefore two lattices do not coincide in general. 
For instance, in $\rSU(N)$ the fundamental weight vectors are not 
parallel to any of the root vectors so that the integral lattice is 
only a sublattice of the central lattice. 

\begin{figure}
\begin{center}
\begin{minipage}[t]{6cm}
\centerline{\hbox{\psfig{file=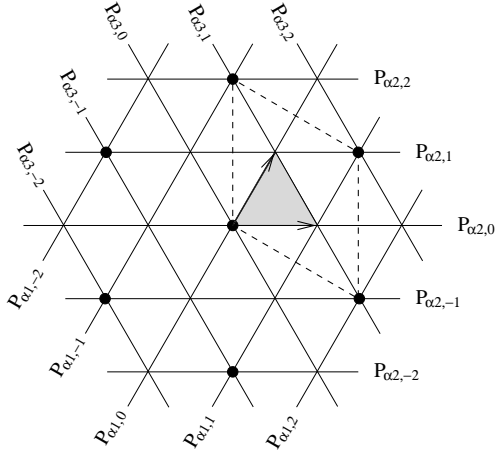,height=6cm}}}
\end{minipage}
\caption{A Weyl domain for $\rSU(3)$ is shaded region. 
All the intersection points consist the central lattice, whereas 
the integral lattice points are indicated by black dots. For $\rSU(3)$, 
the integral lattice coincides with the co-root lattice. 
The parallelogram of dashed lines provides a unit cell of the integer lattice.}
\label{fig:su3w}
\end{center}
\end{figure}

For both simple and non-simple roots, $P_{\alpha_i,\,n_i}$ is 
perpendicular to the corresponding root vector $\vec{\alpha}_i$. 
A mirror reflection on $P_{\alpha_i,\,n_i}$ is therefore nothing but 
the action of the extended Weyl group, which is the semi-direct product of 
Weyl group and the translation by the co-root lattice.\footnote{
The co-root of a root $\alpha$ can be defined as $2\alpha/|\alpha|^2$ 
and co-root lattice is a lattice generated by co-roots. 
In all simply-laced cases, co-roots and roots are identical since 
there are only long roots with length $\sqrt2$.} 
We call a minimal region surrounded by all possible $P_{\alpha_i,\,n_i}$'s 
a Weyl domain. Then the fundamental domain mentioned before turns out to be 
a Weyl domain, a fundamental domain of the extended Weyl group. 
The problem of symmetry 
enhancement therefore reduces to classifying the intersections of 
$P_{\alpha_i,\,n_i}$'s. The dimension of a D-brane is determined 
according to its position $\vec{\psi}$ in the weight space. 
Although the moduli space of D-branes is naturally identified 
with the maximal torus, we may consider only the one of Weyl domains in 
the weight space. Any element of a unit cell can be mapped to a 
point in the Weyl domain by the reflection about the hyperplanes. 
In figure \ref{fig:su3w}, we draw the Weyl domain and lattices mentioned 
above for SU(3). 
The integral lattice points are indicated by black dots, while the central 
lattice points are all the intersection points. 
The parallelogram of dashed lines provides a unit cell of the integer lattice.

\begin{figure}
\begin{center}
\begin{minipage}[t]{5cm}
\centerline{\hbox{\psfig{file=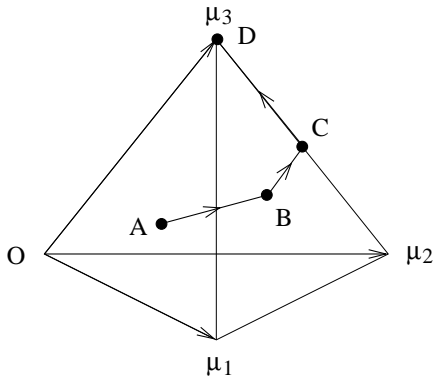,height=5cm}}}
\end{minipage}
\caption{Centralizer enhancement in $G=\rSU(4)$. As $X$ moves to higher 
co-dimensional boundaries, the centralizer is enhanced so that at the four 
vertices the centralizer is $\rSU(4)$ itself. At the points A, B, C, D, 
the centralizer is given by $\rU(1)^3$, $\rU(1)^2\times\rSU(2)$, 
$\rU(1)\times\rSU(3)$, $\rSU(4)$, respectively.}
\label{fig:tetra}
\end{center}
\end{figure}

Let us demonstrate how the centralizer is enhanced depending on the 
location in the weight space. Recall that any point $h$ on the maximal 
torus $T$ is given by $h \equiv e^{2\pi i X}$. 
Start from a generic point $X$ within a Weyl domain whose centralizer is 
the maximal torus itself, namely $\rU(1)^r$. 
Corresponding D-brane has the dimension $p=\dim G -{\rm rank}\,G$. 
Now we move $X$ to one of the hyperplanes $P_{\alpha_i,n}$. 
This hyperplane is co-dimension 1 boundary of the Weyl domain.
Then $\cZ(h)$ is enhanced to 
$\rU(1)^{r-1}\times\rSU(2)_{\alpha_i}$. 
If $X$ further moves to an intersection of two hyperplanes, say 
$P_{\alpha_i,\,n_i}$ and $P_{\alpha_j,\,n_j}$, 
then $\cZ(h)$ is enhanced to 
$\rU(1)^{r-2}\times\rSU(2)_{\alpha_i}\times\rSU(2)_{\alpha_j}$. 
Moreover, if the corresponding root vectors $\vec{\alpha}_i$ and 
$\vec{\alpha}_j$ are not orthogonal to each other, 
the $\rSU(2)\times\rSU(2)$ is further enhanced to $\rSU(3)$ so that the 
centralizer $\cZ(h)$ is further enlarged to $\rU(1)^{r-2}\times\rSU(3)$. 
This process can be continued until we arrive at a central 
lattice point whose centralizer is the whole group, corresponding 
to a D0-brane. In figure \ref{fig:tetra}, we draw a Weyl domain and 
the centralizer enhancement for SU(4).

\begin{figure}
\begin{center}
\begin{minipage}[t]{3.5cm}
\centerline{\hbox{\psfig{file=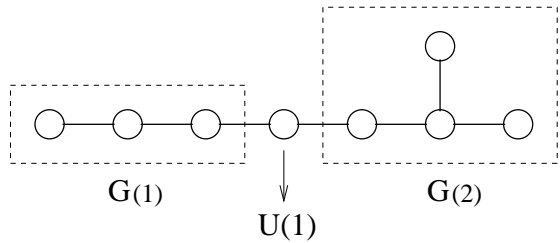,height=3.5cm}}}
\end{minipage}
\caption{The method of iterative deletion. The first step is illustrated 
for $G=D_8$ and when $\psi_4$ is an non-integer. Dynkin diagram is divided 
into 2 disjoint sub-diagrams ($\kappa=2$), and the deleted blob corresponds 
to U(1) that is not enhanced. After the first step, the centralizer is given 
by $\rU(1)\times G_{(1)} \times G_{(2)}$ with $G_{(1)}=A_3$, $G_{(2)}=D_4$.} 
\label{fig:dynkin}
\end{center}
\end{figure}

The general recipe of the D-brane classification can be described 
in terms of Dynkin diagram as follows. 
Suppose a D-brane location specified by $\vec{\psi}$ belongs to the 
intersection of $k$ hyperplanes.\footnote{
The reader should not confuse this $k$ with the level number $k$ 
of WZW model.} 
Out of the Dynkin diagram of $G$, we take away $r-k$ roots (blobs) 
corresponding to U(1)'s which are not enhanced to SU(2)'s 
(See figure \ref{fig:dynkin}). 
Then the Dynkin diagram becomes a disjoint union of, say $\kappa$, 
sub-diagrams, each of which corresponds to a subgroup $G_{(i)}$ 
of the original group $G$. We allow $G_{(i)}$'s to be trivial, namely 
$G_{(i)}$'s can be the trivial group that consists of only the identity 
element. If any of $G_{(i)}$'s is trivial 
the corresponding sub-diagram is an empty box without any blob.
Then the centralizer at $\vec{\psi}$ is given by 
\be
\cZ(\vec{\psi})=\rU(1)^{r-k}\times\prod_{i=1}^\kappa G_{(i)}, \label{Zh}
\ee 
where $r\equiv{\rm rank}\,G$ as before. 
Define $r_i \equiv {\rm rank}\, G_{(i)}$ allowing $r_i=0$ which means 
that $G_{(i)}$ is trivial. The possible subgroups $G_{(i)} \subset G$ 
must obey the constraint:
\be
\sum_{i=1}^\kappa r_i=k. \label{rankGi}
\ee
The number of sub-diagrams $\kappa$ depends on the topology of Dynkin diagram 
of $G$. The corresponding D-brane $D(\vec{\psi})$ is given by the coset 
\be
D(\vec{\psi})=\frac{G}{\rU(1)^{r-k}\times\prod_{i=1}^{\kappa} G_{(i)}}
\quad{\rm if}\quad 
\vec{\psi} \in \bigcap_{i=1}^k P_{\alpha_i,n_i}
\ee
which determines the dimension of the D-brane as 
\be
p=\dim G -(r-k) -\sum_{i=1}^{\kappa} \dim G_{(i)}. \label{Ddim}
\ee
 
In order to find out all the possible D-branes, one can use the method of 
iterative deletion. Namely, we start from the Dynkin diagram of a given 
group $G$. Deleting a blob divides the diagram into two pieces and we get 
a centralizer that is the product of three factors: namely, 
$G_{(1)}$, $G_{(2)}$, and the U(1) that comes from the deleted blob. 
According to the position of the deleted blob, there are $r$ possible results 
in the first step for rank $r$ group. One can iterate this procedure by 
applying it to any of the factor groups $G_{(i)}$'s.
If the end blob is deleted, one can consider one of $G_{(i)}$'s is an 
empty set.
In figure \ref{fig:dynkin}, we demonstrated this method for $G=D_8$. 
Since each blob corresponds to a simple root, this method takes care of 
(intersections of) hyperplanes $P_{\alpha_i,n_i}$ only for the simple 
roots $\alpha_i$'s. However, for any set of linearly independent roots, 
one can form a simple root system including the given roots by changing 
the signs of some of them. Therefore it is enough to consider the simple 
roots only and the Dynkin diagram method do not miss any type of D-branes.

In the following, we will apply the above recipe to the simply-laced 
Lie groups $G=A_r$, $D_r$, $E_6$, $E_7$, $E_8$. 
For non-simply-laced cases, we discuss in the separate paper \cite{IS2}. 
First let us look at the centralizer enhancement in $A_r$-series 
($A_r=\rSU(r+1)$). The Dynkin diagram for $A_r$ has a topology 
of a line segment without branching so that $\kappa=r-k+1$ and 
$G_{(i)}=A_{r_i}$ defining $A_0$ as the trivial group. 
This can be understood since, in the Dynkin diagram of $A_r$, deleting $r-k$ 
blobs of unenhancement results in the disjoint $r-k+1$ connected pieces 
corresponding to $A_{r_i}$'s. 
The general formula (\ref{Zh}) of $\cZ$ now becomes
\be
\cZ(\vec{\psi})=\rU(1)^{r-k}\times\prod_{i=1}^{r-k+1} A_{r_i},
\ee
with the condition
\be
\sum_{i=1}^{r-k+1} r_i=k. \label{rankAr}
\ee
The dimension of the D-brane is given by 
\be
p=r(r+1)-\sum_{i=1}^{r-k+1} r_i(r_i+1),
\ee
where we have used Eq.\ (\ref{rankAr}) to cancel $k$ in Eq.\ (\ref{Ddim}).
Notice that the dimension $p$ always becomes an even number. 
This unexpected result can be understood if one consider the equivalence 
of the adjoint orbit and the co-adjoint orbit for any compact Lie group 
and the fact that the coadjoint orbit admits the symplectic structure, 
which is possible only if the dimension is even. 
This result holds for any simply-laced Lie group as we will see 
explicitly for both $D_r$- and $E_r$-series shortly. 

The D-branes in $A_r$-series are summarized in table 1.
In the table the lower bound of rank $r$ is given for each enhancement 
pattern. Denoting the number of $A_p$'s $(p\ge 1)$ contained in a given 
enhancement pattern by $l_p$, the lower bound of $r$ is obtained as
\be
r \ge \sum_{p=1}^r l_p +k-1.
\ee
This is because, in the Dynkin diagram of $A_r$, $A_p$'s $(p\ge 1)$ 
in a given enhancement pattern should be disjoint blocks remaining 
after deleting $r-k$ blobs that correspond to the unenhanced U(1)'s. 
In order to get the total 
$L\equiv\sum_{p=1}^r l_p$ disjoint blocks, we have to delete 
at least $L-1$ blobs from the Dynkin diagram so that $r-k\ge L-1$ 
to yield the lower bound.

\begin{table}
\begin{tabular}{|c|l|r|l|l|}\hline
  $k$ & centralizer Z & dim (Z) & dim.\ of D-brane & comments \\ \hline\hline 
    0 & $\rU(1)^{r}$ & $r$ & $r(r+1)$ & generic \\ \hline 
    1 & $\rU(1)^{r-1}\times A_1$ & $3+r-1$ & $r(r+1)-2$ & $r\geq 1$ \\ \hline 
    2 & $\rU(1)^{r-2}\times A_1^{2}$ & $3\times 2+r-2$ & $r(r+1)-4$ & 
        $r\geq 3$ \\ \hline 
    2 & $\rU(1)^{r-2}\times A_2$ & $8+r-2$ & $r(r+1)-6$ & $r\geq 2$ \\ \hline
    3 & $\rU(1)^{r-3}\times A_1^{3}$ & $3\times 3+r-3$ & $r(r+1)-6$ & 
        $r\geq 5$ \\ \hline
    3 & $\rU(1)^{r-3}\times A_1\times A_2$ & $3+8+r-3$ & $r(r+1)-8$ & 
        $r\geq 4$ \\ \hline
    3 & $\rU(1)^{r-3}\times A_3$ & $15+r-3$ & $r(r+1)-12$ & 
        $r\geq 3$ \\ \hline 
    4 & $\rU(1)^{r-4}\times A_1^{4}$ & $3\times 4+r-4$ & $r(r+1)-8$ & 
        $r\geq 7$ \\ \hline 
    4 & $\rU(1)^{r-4}\times A_1^{2}\times A_2$ & $3\times 2+8+r-4$ & 
        $r(r+1)-10$ & $r\geq 6$ \\ \hline 
    4 & $\rU(1)^{r-4}\times A_2^{2}$ & $8\times 2+r-4$ & $r(r+1)-12$ & 
    $r\geq 5$ \\ \hline 
    4 & $\rU(1)^{r-4}\times A_1\times A_3$ & $3+15+r-4$ & $r(r+1)-14$ & 
    $r\geq 5$ \\ \hline
    4 & $\rU(1)^{r-4}\times A_4$ & $24+r-4$ & $r(r+1)-20$ &  
    $r\geq 4$ \\ \hline
      & $\cdots$ & $\cdots$ & $\cdots$ & \\ \hline
$r-1$ & $\rU(1)\times A_{r_1}\times A_{r_2}$ & $\sum_i r_i(r_i+2)+1$ & 
    $r(r+1)- \sum_i r_i(r_i+1)$ & $r=r_1+r_2+1$ \\ \hline
  $r$ & $A_r$ & $r(r+2)$ & 0 & D0 \\ \hline
\end{tabular}
  \centering
  \caption{D-branes in $A_r =\rSU(r+1)$.}\label{table:su}
\end{table}

For $D_r$-series ($D_r=\rSO(2r)$ with $r\ge 2$), D-branes and their 
dimensions can be worked out as follows:
For $k=0,1,2$ with sufficiently large rank, the centralizer structure 
is the same as in $A_r$-series. 
So one can refer the table 1, except for the shift 
$r(r+1) \to 2r(r-1)$ in the D-brane dimension $p$.
For $k=3$ and higher, the general structure of D-branes can be worked out 
inductively by descending $D_r$-series. 
So we start from $D_4=\rSO(8)$ ($r=4$), that is the first non-trivial 
element in $D_r$-series. For $k=0,1,2$ we have already discussed above. 
For $k=3$ we can have either $\rU(1)\times D_3$ or $\rU(1)\times A_1^3$, 
depending on which root is deleted. 
The corresponding D-branes have dimensions 12 and 18 respectively. 
The case of $k=4$  simply gives D0-branes.

Now we work on $D_r$ with $r\geq 5$ by mathematical induction.
Suppose we know how to describe D-branes in $D_{r_1}$ $(r_1 \le r-1)$ 
and in $A_{r_2}$ $(r_2 \le r-1)$. 
Setting $k=r$ in $D_r$, the centralizer is $D_r$ itself and 
the corresponding D-brane is D0. For $k=r-1$, possible patterns are 
$\rU(1)\times D_{r_1}\times A_{r_2}$ with the constraint $r_1+r_2=r-1$, 
$r_1\neq 1$. 
To get the patterns for $k=r-2$, we only have to repeat the same procedure 
in either $D_{r_1}$ or $A_{r_2}$ which arise in the patterns for $k=r-1$.
Since we know all the patterns for $A_{r_1}$ and for $D_{r_2}$ by induction 
hypothesis, we finish the job. The patterns for $k=r-2$ are given by 
$\rU(1)^{2}\times D_{r_1}\times A_{r_2}\times A_{r_3}$ with 
$r_1+r_2+r_3=r-2$, $r_1\neq 1$. 
One can go down to lower $k$'s by iterating the procedure 
without any new feature. 
For the general $k$, the centralizer is given by 
 \be 
 \cZ=\rU(1)^{r-k} \times D_{r_1} \times \prod_{i=2}^{r-k+1}A_{r_i}.
 \ee
The dimension of a D-brane in $D_r$-series is given by 
\bea
p&=&r(2r-1)-(r-k)-r_1(2 r_1-1)-\sum_{i=2}^{r-k+1} r_i(r_i+2)
\nonumber \\
&=& 2r(r-1)-2r_1(r_1-1)-\sum_{i=2}^{r-k+1}r_i(r_i+1).
\eea
  Notice that it is manifestly an even integer as we discussed before.
\begin{table}
  \begin{tabular}{|c|l|r|c|} \hline
 k & centralizer Z & dim (Z) & dim.\ of D-brane  \\ \hline\hline
 5 & $\rU(1)\times D_5$  & 1+45=46 & 32  \\ \hline
 5 &$\rU(1)\times A_5$  &1+35=36 & 42 \\ \hline
 5 & $\rU(1)\times A_1\times A_4$ & 1+3+24=28 & 50  \\ \hline
 5 & $\rU(1)\times A_1\times A_2^2$  &1+3+8+8=20 &58 \\ \hline
\end{tabular}
 \centering
  \caption{D-branes in $E_6$ ($\dim E_6=78$). We described only $k=5$. 
   For a given pattern of the centralizer, one can go down to lower $k$'s 
   by deleting one of the SU(2) blobs in one of the non-abelian factors.
}\label{e6}
\vspace*{36pt}
  \begin{tabular}{|c|l|r|c|} \hline
 k & centralizer Z & dim (Z) & dim.\ of D-brane  \\ \hline\hline
 6 & $\rU(1)\times E_6$ & 1+78=79 & 54  \\ \hline 
 6 & $\rU(1)\times D_6$ & 1+66=67 & 66  \\ \hline
 6 & $\rU(1)\times A_1\times D_5$ & 1+3+45=49 & 84  \\ \hline
 6 & $\rU(1)\times A_1\times A_5$ & 1+3+35=39 & 94  \\ \hline
 6 & $\rU(1)\times A_2\times A_4$ & 1+8+24=33 & 100  \\ \hline
 6 & $\rU(1)\times A_1\times A_2\times A_3$ & 1+3+8+15=27 &106  \\ \hline
\end{tabular}
 \centering
  \caption{D-branes in $E_7$ ($\dim E_7=133$). We described only $k=6$.
   One can continue to lower $k$ by iteration.}\label{e7}
\vspace*{36pt}
  \begin{tabular}{|c|l|r|c|} \hline
 k & centralizer Z & dim (Z) & dim.\ of D-brane  \\ \hline\hline
 7 & $\rU(1)\times E_7$ & 1+133=134 & 114  \\ \hline
 7 & $\rU(1)\times D_7$ & 1+91=92 & 156  \\ \hline
 7 & $\rU(1)\times A_1\times E_6$ & 1+3+78=82 & 166  \\ \hline
 7 & $\rU(1)\times A_7$ & 1+63=64 & 184  \\ \hline
 7 & $\rU(1)\times A_2\times D_5$ & 1+8+45=54 & 194  \\ \hline
 7 & $\rU(1)\times A_1\times A_6$ & 1+3+48=52 & 196  \\ \hline
 7 & $\rU(1)\times A_3\times A_4$ & 1+15+24=40 & 208  \\ \hline
 7 & $\rU(1)\times A_1\times A_2\times A_4$ & 1+3+8+24=36 & 212  \\ \hline
\end{tabular}
 \centering
  \caption{D-branes in $E_8$ ($\dim E_8=248$). We described only $k=7$.
   One can continue to lower $k$ by iteration.}\label{e8}
\end{table}

Similarly it is enough to show the patterns for $k=r-1$ in $E_r$-series 
$r=6$, $7$, $8$. For a given pattern of the centralizer, 
one can go down to lower $k$'s by deleting one of the SU(2) blobs 
in one of the non-abelian factors. 
We tabulate the results in tables \ref{e6}, \ref{e7}, \ref{e8}.

\setcounter{equation}{0}
\section{Quantum stability and Flux of WZW D-branes}

In earlier section, we have seen that 
the single-valuedness of the path integral of level $k$ WZW action with 
boundary gives two quantization conditions: one from the $H$-monopole 
 and the other from the $F$-monopole. The former condition gives 
the quantization of the level $k$, while the 
latter gives the condition that D$(p-2)$-brane charge is integer modulo $k$.
In this section we determine the charge in terms of the 
parameters that describe the position of a D-brane.
During that process we will see that the nature of the D-brane charge 
is in fact not a scalar, but a vector, which is nothing but the discretized 
$\vec{\psi}$ vector itself.
The Wilson loop expectation value, is therefore a measure of location of the 
D-brane, which is a familiar result in the flat space D-brane theory.

For $G={\rm SU(2)}$, it was found in \cite{BDS} that the DBI action of 
a spherical D2-brane is minimized when the $F$-monopole flux of the D2-brane 
coincides with its transverse U(1) coordinate $\psi$ multiplied by $2\pi k$. 
The quantization of the D0-charge therefore protects the D2-brane from 
degenerating to a point. Via the quantization of $F$-monopole 
flux, the transverse U(1) coordinate of the D2-brane is discretized.
The D2-brane is thereby fixed at a discrete point $\psi=n/k$, 
where it is stable. 

We now want to generalize the aforementioned fact to higher 
dimensional groups. The foliation $G/T \times T $ implies that 
the transverse coordinates of D-branes are homology 1-cycles of $T $, 
which are generated by $H_\alpha$'s. For a compact Lie group $G$, 
one can therefore apply the stabilization mechanism to each $\rSU(2)_\alpha$ 
subgroup of $G$. For more precise arguments, we consider a 2-cycle $S$ 
embedded in a D-brane and suppose $S$ wraps the base homology 2-cycles 
$\{S_\alpha=\rSU(2)_\alpha/\rU(1)_\alpha\}$ such that 
\be
S \cong\sum_\alpha\, c_\alpha S_\alpha, \label{decomS}
\ee
whose precise meaning is given by 
\be
\int_S F =\sum_\alpha \,c_\alpha\!\int_{S_\alpha}\!F. \label{decomF}
\ee
Each integral coefficient $c_\alpha$ represents the winding 
number of $S$ over $S_\alpha$. 
Notice that since $F$ is a closed form, smooth deformation of the 
base 2-cycle $S$ does not change the value of $\int_SF$. 
Now recall that $\int_S F$ is $2\pi$ times an integer and this is 
true for any choice of the surface $S$, i.e, for any choice of the set 
${c_\alpha}$. 
By choosing the vector $(c_1,c_2, \cdots, c_r)$ as a unit vector 
$(1,0,\cdots,0)$, etc, we see that 
\be
\int_{S_\alpha} F=2\pi n_\alpha \quad {\rm mod~} 2\pi k
~~{\rm for~all~~} \alpha \label{na}.
\ee 

On the other hand by an explicit calculation using 
Eqs.\ (\ref{H}), (\ref{giF}), one gets \cite{G} 
\be
\int_{S_\alpha} F=\int_{S_\alpha} \cF- \int_{Z_\alpha} H 
= 2\pi k\,\vec{\alpha}\cdot\vec{\psi} \label{psia}.
\ee
Comparing Eq.\ (\ref{na}) with Eq.\ (\ref{psia}), we get 
\be
\vec{\alpha}\cdot\vec{\psi}=\frac{n_\alpha}{k} . \label{5.5}
\ee
The meaning of this is deep: For the quantum mechanical consistency, 
the position of a D-brane given by $\vec{\psi}$ should be quantized.
Classically, the singular D-branes are of measure 0 in the D-brane moduli 
space. Owing to the discretization of the moduli space, the singular 
D-branes now take finite fraction of it. 
Notice that $2\pi k$ modularity of the $F$-monopole charge is nothing but 
the periodicity of enhancement in the weight space.
We have seen that $k\,\vec{\alpha_i}\cdot\vec{\psi}$ should be
an integer for all positive roots $\alpha_i$. This can be equivalently 
stated that $k\,\vec{\psi}$ should be a highest weight of an irreducible 
representation. 
 
One can understand the vectorial nature of the D$(p-2)$-charge $\int F$ 
by using the cohomological consideration. 
As is shown in appendix A, we have 
\be
H^2(Q;\Z)\cong\pi_1(T ). \label{iso}
\ee
The right hand side is trivially given by ${\bf Z}^r$ whose element can be 
identified with ($\psi_1$, $\psi_2$, $\cdots$, $\psi_r$) where 
$\psi_i=\vec{\psi}\cdot\alpha_i$, showing that the homology elements 
(or $F$-monopole charges) are in one to one correspondence to 
the discretized position vectors of the D-brane. 

So far our discussion for the quantization is purely topological. 
However one can show that the DBI action of a D-brane is minimized 
only when the D-brane is located within the finite lattice as shown 
explicitly in SU(2) case by \cite{BDS}. Since the calculation in the general 
group $G$ is the same as in SU(2), we do not repeat it here. 
Final comment is that the level $k$ 
should be shifted to $k+N$ for $\rSU(N)$ for example so that 
Eq.\ (\ref{5.5}) should be changed to 
\be
 \vec{\alpha}\cdot\vec{\psi}=\frac{n_\alpha}{k+N} .
\ee
For the general group $G$, $N$ should be replaced by the dual coxeter 
number of $G$. For SO($N$), it is $N-2$, and for Sp($N$), it is $N+1$.
This shift is well known in CFT and is due to the quantum corrections.

\setcounter{equation}{0}
\section{Conclusion}

In this paper, we gave a simple review of D-branes in group 
manifolds and the classification of the singular D-branes according to 
their locations in the fundamental domain in the weight space. 
The general recipe has a simple description in terms of Dynkin diagram. 
The dimension of a D-brane is always given by an even number 
for untwisted D-branes and reduces as we go from a generic point of 
the fundamental domain to its higher co-dimensional boundaries. 
We also described how the positions of D-branes in the fundamental domain 
are restricted to a discrete subset corresponding to the highest weight 
irreducible representations. 

We point out some of the future works.
First of all our discussion is confined to the untwisted D-branes. 
The twisted case involves outer automorphisms that preserve the topology 
of Dynkin diagram, and it can change the dimension of a D-brane 
\cite{FrSc,S3}. 
Consideration of the full classification of outer automorphisms 
is doable and possible future problem. 
Second we discussed only simply-laced cases, where only long roots are 
involved. Including both long and short roots introduces new issues where 
the shape of fundamental Weyl domain is more involved and the global issue 
becomes non-trivial. 
We will discuss this problem in a separate paper \cite{IS2}. 
Other interesting problem is generalizing the problem to the coset case, 
which was in fact the goal of this project. 

Finally, our discussion of 
$F$-monopole charges as D$(p-2)$-brane charges is based on (co)homology 
groups and is not accurate. As pointed out in \cite{W2}, K-theory groups 
(or twisted K-theory groups in the presence of nontrivial $H$) are natural 
to argue D-brane charges. Recent arguments in \cite{FrSc,MMS} have shown 
that the twisted K-theory groups capture the correct modulus of monopole flux 
$F$ ($k\to k+N$ for $\rSU(N)$). Another interesting approach is a quantum 
algebraic description of WZW D-branes proposed in \cite{PS} where the 
$q$-deformation of the Lie algebra $\mg$ plays an essential role to show 
the quantum stability of singular branes discussed in section 5. 
We postpone more on the K-theoretic or quantum algebraic studies 
of WZW D-branes to future work.

\vspace{18pt}
 
\noindent
{\bf Acknowledgements}\\
This work is supported by KOSEF 1999-2-112-003-5.

\appendix

\renewcommand{\thesection}{\large \bf \mbox{Appendix~}\Alph{section}}
\renewcommand{\theequation}{\Alph{section}\mbox{.}\arabic{equation}}

\setcounter{equation}{0} 
\section{Proof of a homological theorem }

We will prove the isomorphic relation $H^2(Q;\Z)\cong \pi_1(T)$
by following Ref. \cite{BDS}. Here we assume that the group $G$ is 
simply-connected. 
One can use the exact homotopy sequence \cite{NS} 
\be
\dots\to\pi_p (G)\to\pi_p (G/T )\to\pi_{p-1} (T )
\to\pi_{p-1} (G)\to\dots \label{homoseq}
\ee
together with the basic fact \cite{NS}:
\be
\{0\}\to A \to B \to \{0\} \quad\Longrightarrow\quad A\cong B. \label{theo}
\ee
Suppose, for a given group space $G$, 
its homotopy groups $\pi_0 (G)$, $\pi_1 (G)$, $\pi_2 (G)$ are all trivial.
Setting $p=1$ in the exact sequence (\ref{homoseq}), one can immediately 
see $\pi_1 (Q)=\{0\}$ which ensures that the D-brane $Q$ is simply-connected. 
Setting $p=2$ provides the non-trivial relation 
$\pi_2 (Q)\cong\pi_1 (T )$, which at the same time means that 
$\pi_2(Q)$ is the first non-trivial homotopy group over the D-brane. 
Since the first non-trivial homology group and the first non-trivial 
homotopy group have the same dimension and are isomorphic \cite{NS}, 
$H^2 (Q;\Z)\cong\pi_2 (Q)$ and we arrive at the desired result. 
If $G$ is not simply-connected, $\pi_1 (G)$ is nontrivial. However, 
we can use the covering group $\widetilde{G}$ of $G$ instead of $G$ itself. 
Then the above proof does work for $\widetilde{G}$ and its maximal torus 
$\widetilde{T}$. 

\setcounter{equation}{0}
\section{Generators in the fundamental representation}

Here we include the explicit forms of generators in the fundamental 
representations for a few groups which are essential for the study of 
singular D-brane classification. Since SU(2) case can be found easily 
elsewhere, we include only SU(3), SO(4), SO(5) and $G_2$. 

\subsection{\large \bf SU(3)}

By using Gell-Mann matrices $\lambda_a$'s shown in \cite{Geo}, 
the Cartan generators of $\mathfrak{su}(3)$ are given by
\be
H_1 \equiv\frac{1}{2}\,\lambda_3 =\frac{1}{2}\,{\rm diag}(1,-1,0),\quad
H_2 \equiv\frac{1}{2}\,\lambda_8 =\frac{1}{2\sqrt{3}}\,{\rm diag}(1,1,-2).
\ee
Then the raising and lowering operators of $\mathfrak{su}(2)_{\alpha_i}$'s 
are determined as
\be
E_{\pm \alpha_1} \equiv \frac{1}{2}\,(\lambda_4 \pm i\lambda_5),\quad 
E_{\pm \alpha_2} \equiv \frac{1}{2}\,(\lambda_6 \mp i\lambda_7),\quad
E_{\pm \alpha_3} \equiv \frac{1}{2}\,(\lambda_1 \pm i\lambda_2).
\ee
The $(J^3, J^\pm)$ operators, namely $(H_{\alpha_i},e_{\pm\alpha_i})$'s 
properly normalized in Eq.\ (\ref{norm}), are given by the following 
3$\times$3 matrices
\bea
&&
H_{\alpha_1} ~=~
\frac{1}{2}\left[
\begin{array}{ccc}
  1 &    &    \\
    &  0 &    \\
    &    & -1 
\end{array}
\right], \quad 
e_{+\alpha_1} ~=~
\frac{1}{\sqrt{2}}\left[
\begin{array}{ccc}
  0 &  0 &  1 \\
  0 &  0 &  0 \\
  0 &  0 &  0 
\end{array}
\right], \label{su3a1} \\
&&
H_{\alpha_2} ~=~
\frac{1}{2}\left[
\begin{array}{ccc}
  0 &    &    \\
    & -1 &    \\
    &    &  1 
\end{array}
\right], \quad 
e_{+\alpha_2} ~=~
\frac{1}{\sqrt{2}}\left[
\begin{array}{ccc}
  0 &  0 &  0 \\
  0 &  0 &  0 \\
  0 &  1 &  0 
\end{array}
\right], \label{su3a2} \\
&&
H_{\alpha_3} ~=~
\frac{1}{2}\left[
\begin{array}{ccc}
  1 &    &    \\
    & -1 &    \\
    &    &  0 
\end{array}
\right], \quad 
e_{+\alpha_3} ~=~
\frac{1}{\sqrt{2}}\left[
\begin{array}{ccc}
  0 &  1 &  0 \\
  0 &  0 &  0 \\
  0 &  0 &  0 
\end{array}
\right], \label{su3a3}
\eea
and their hermitian conjugations. 

\subsection{\large \bf SO(4)}

Introducing the $\rSO(4)$ angular momentum operators $M_{ab}$, 
the Cartan generators are given by $(H_1,H_2)=(M_{12},M_{34})$. 
The raising and lowering operators of $\mathfrak{su}(2)_{\alpha_i}$'s 
are given by
\bea
&&
E_{+\alpha_1} \equiv \frac{1}{2}(M_{24}+M_{13}+iM_{23}-iM_{14}),\nonumber\\
&&
E_{+\alpha_2} \equiv \frac{1}{2}(M_{24}-M_{13}-iM_{23}-iM_{14}),
\eea
and their hermitian conjugations.
In the fundamental representation of $\rSO(4)$, $M_{ab}$'s are given by 
the 4$\times$4 matrices 
$(M_{ab})_{cd} \equiv -i(\delta_{ac}\delta_{bd}-\delta_{ad}\delta_{bc})$.
Diagonalizing the above generators by using a unitary matrix
\be
U =\frac{1}{\sqrt{2}}\left[
\begin{array}{ccccc}
 0 &  0 &  1 & -i \\
-i &  1 &  0 &  0 \\
 i &  1 &  0 &  0 \\
 0 &  0 &  1 &  i 
\end{array}
\right],
\ee
we obtain a representation where the Cartan generators 
$(H_1,H_2)=U(M_{12},M_{34})\,U^\dagger$ 
are given by the 4$\times$4 diagonal matrices
\be
H_1 ~=~{\rm diag}(0,-1,1,0), \quad H_2 ~=~{\rm diag}(1,0,0,-1).
 \label{so4cartan}
\ee
The $(J^3, J^\pm)$ operators are then given by the following 
4$\times$4 matrices
\bea
&&
H_{\alpha_1} ~=~
\frac{1}{2}\left[
\begin{array}{cccc}
 -1 &    &    &    \\
    & -1 &    &    \\
    &    &  1 &    \\
    &    &    &  1 
\end{array}
\right], \quad 
e_{+\alpha_1} ~=~
\frac{1}{\sqrt{2}}\left[
\begin{array}{cccc}
  0 &  0 &  0 &  0 \\
  0 &  0 &  0 &  0 \\
  1 &  0 &  0 &  0 \\
  0 & -1 &  0 &  0 
\end{array}
\right], \label{so4a1} \\
&&
H_{\alpha_2} ~=~
\frac{1}{2}\left[
\begin{array}{cccc}
  1 &    &    &    \\
    & -1 &    &    \\
    &    &  1 &    \\
    &    &    & -1 
\end{array}
\right], \quad 
e_{+\alpha_2} ~=~
\frac{1}{\sqrt{2}}\left[
\begin{array}{cccc}
  0 &  1 &  0 &  0 \\
  0 &  0 &  0 &  0 \\
  0 &  0 &  0 & -1 \\
  0 &  0 &  0 &  0 
\end{array}
\right], \label{so4a2}
\eea
and their hermitian conjugations.

\subsection{\large \bf SO(5)}

The Cartan generators of SO(5) are $(H_1,H_2)=(M_{12},M_{34})$ by using 
$\rSO(5)$ angular momentum operators $M_{ab}$. 
Introducing raising and lowering operators along $H_1$ and $H_2$ directions:
\be
E_{\pm e_1} \equiv \frac{1}{\sqrt{2}}\left(M_{15} \pm i M_{25}\right),\quad
E_{\pm e_2} \equiv \frac{1}{\sqrt{2}}\left(M_{35} \pm i M_{45}\right),
\label{Ee}
\ee
the raising and lowering operators of $\mathfrak{su}(2)_{\alpha_i}$'s 
are given by
\bea
&& E_{+\alpha_1} \equiv -i[E_{+e_1},E_{-e_2}],\quad
   E_{+\alpha_2} \equiv E_{+e_2},\nonumber\\
&& E_{+\alpha_3} \equiv -i[E_{+e_1},E_{+e_2}],\quad
   E_{+\alpha_4} \equiv E_{+e_1},
\label{Eal}
\eea
and their hermitian conjugations. In the SO(5) fundamental representation, 
$(M_{ab})_{cd} \equiv -i(\delta_{ac}\delta_{bd}-\delta_{ad}\delta_{bc})$. 
Diagonalizing the above generators by using a unitary matrix
\be
U_1 =\frac{1}{\sqrt{2}}\left[
\begin{array}{ccccc}
 0 &  0 &  1 & -i &  0 \\
 1 & -i &  0 &  0 &  0 \\
 0 &  0 &  0 &  0 &  1 \\
 1 &  i &  0 &  0 &  0 \\
 0 &  0 &  1 &  i &  0
\end{array}
\right],
\ee
we obtain a representation where the Cartan generators 
$(H_1,H_2)=U_1(M_{12},M_{34})\,U_1^\dagger$ 
are given by the 5$\times$5 diagonal matrices
\be
H_1 ~=~{\rm diag}(0,1,0,-1,0), \quad H_2 ~=~{\rm diag}(1,0,0,0,-1).
 \label{cartan}
\ee
The $(J^3, J^\pm)$ operators are then given by the 5$\times$5 matrices
\bea
&&
H_{\alpha_1} ~=~
\frac{1}{2}\left[
\begin{array}{ccccc}
 -1 &    &    &    &    \\
    &  1 &    &    &    \\
    &    &  0 &    &    \\
    &    &    & -1 &    \\
    &    &    &    &  1
\end{array}
\right], \quad 
e_{+\alpha_1} ~=~
\frac{1}{\sqrt{2}}\left[
\begin{array}{ccccc}
  0 &  0 &  0 &  0 &  0 \\
 -i &  0 &  0 &  0 &  0 \\
  0 &  0 &  0 &  0 &  0 \\
  0 &  0 &  0 &  0 &  0 \\
  0 &  0 &  0 &  i &  0
\end{array}
\right], \label{so5a1} \\
&&
H_{\alpha_3} ~=~
\frac{1}{2}\left[
\begin{array}{ccccc}
  1 &    &    &    &    \\
    &  1 &    &    &    \\
    &    &  0 &    &    \\
    &    &    & -1 &    \\
    &    &    &    & -1
\end{array}
\right], \quad 
e_{+\alpha_3} ~=~
\frac{1}{\sqrt{2}}\left[
\begin{array}{ccccc}
  0 &  0 &  0 & -i &  0 \\
  0 &  0 &  0 &  0 &  i \\
  0 &  0 &  0 &  0 &  0 \\
  0 &  0 &  0 &  0 &  0 \\
  0 &  0 &  0 &  0 &  0
\end{array}
\right], \label{so5a3}
\eea
for long roots $\alpha_1$ and $\alpha_3$, while for short roots $\alpha_2$ 
and $\alpha_4$ they are
\bea
&&
H_{\alpha_2} ~=~
\left[
\begin{array}{ccccc}
  1 &    &    &    &    \\
    &  0 &    &    &    \\
    &    &  0 &    &    \\
    &    &    &  0 &    \\
    &    &    &    & -1
\end{array}
\right], \quad  
e_{+\alpha_2} ~=~
\left[
\begin{array}{ccccc}
  0 &  0 & -i &  0 &  0 \\
  0 &  0 &  0 &  0 &  0 \\
  0 &  0 &  0 &  0 &  i \\
  0 &  0 &  0 &  0 &  0 \\
  0 &  0 &  0 &  0 &  0
\end{array}
\right], \label{so5a2} \\
&&
H_{\alpha_4} ~=~
\left[
\begin{array}{ccccc}
  0 &    &    &    &    \\
    &  1 &    &    &    \\
    &    &  0 &    &    \\
    &    &    & -1 &    \\
    &    &    &    &  0
\end{array}
\right], \quad 
e_{+\alpha_4} ~=~
\left[
\begin{array}{ccccc}
  0 &  0 &  0 &  0 &  0 \\
  0 &  0 & -i &  0 &  0 \\
  0 &  0 &  0 &  i &  0 \\
  0 &  0 &  0 &  0 &  0 \\
  0 &  0 &  0 &  0 &  0
\end{array}
\right]. \label{so5a4}
\eea
The lowering operators $e_{-\alpha_i}$'s are given by hermitian 
conjugations of $e_{+\alpha_i}$'s above. 

\subsection{\large \bf Sp(2)}

The fundamental representation of Sp(2) is given by the 4-dimensional spinor 
representation of SO(5). According to the standard construction in \cite{Geo}, 
the SO(5) angular momentum operators are given by
\bea
&&
M_{12}=\frac{1}{2}\,\sigma_3 \otimes I, \quad
M_{34}=\frac{1}{2}\,I \otimes \tau_3, \\
&&
M_{15}=\frac{1}{2}\,\sigma_1 \otimes I, \quad
M_{35}=\frac{1}{2}\,\sigma_3 \otimes \tau_1, \\
&&
M_{25}=\frac{1}{2}\,\sigma_2 \otimes I, \quad
M_{45}=\frac{1}{2}\,\sigma_3 \otimes \tau_2,
\eea
where both $\sigma_1$'s and $\tau_i$'s are Pauli matrices.
The Cartan generators $(H_1,H_2)=(M_{12},M_{34})$ are given by
\be
H_1 ~=~\frac{1}{2}{\rm diag}(1,1,-1,-1), \quad 
H_2 ~=~\frac{1}{2}{\rm diag}(1,-1,1,-1),
 \label{sp2cartan}
\ee
without diagonalization. The raising and lowering operators are constructed 
in the same way as in Eqs.\ (\ref{Ee}), (\ref{Eal}). 
The $(J^3, J^\pm)$ operators are then given by the 4$\times$4 matrices
\bea
&&
H_{\alpha_1} ~=~
\frac{1}{2}\left[
\begin{array}{cccc}
  0 &    &    &    \\
    &  1 &    &    \\
    &    & -1 &    \\
    &    &    &  0 
\end{array}
\right], \quad 
e_{+\alpha_1} ~=~
\frac{1}{\sqrt{2}}\left[
\begin{array}{cccc}
  0 &  0 &  0 &  0 \\
  0 &  0 &  i &  0 \\
  0 &  0 &  0 &  0 \\
  0 &  0 &  0 &  0 
\end{array}
\right], \label{sp2a1} \\
&&
H_{\alpha_3} ~=~
\frac{1}{2}\left[
\begin{array}{cccc}
  1 &    &    &    \\
    &  0 &    &    \\
    &    &  0 &    \\
    &    &    & -1 
\end{array}
\right], \quad 
e_{+\alpha_3} ~=~
\frac{1}{\sqrt{2}}\left[
\begin{array}{cccc}
  0 &  0 &  0 &  i \\
  0 &  0 &  0 &  0 \\
  0 &  0 &  0 &  0 \\
  0 &  0 &  0 &  0 
\end{array}
\right], \label{sp2a3}
\eea
for long roots $\alpha_1$ and $\alpha_3$, while for short roots $\alpha_2$ 
and $\alpha_4$ they are
\bea
&&
H_{\alpha_2} ~=~
\frac{1}{2}\left[
\begin{array}{cccc}
  1 &    &    &    \\
    & -1 &    &    \\
    &    &  1 &    \\
    &    &    & -1 
\end{array}
\right], \quad  
e_{+\alpha_2} ~=~
\frac{1}{\sqrt{2}}\left[
\begin{array}{cccc}
  0 &  1 &  0 &  0 \\
  0 &  0 &  0 &  0 \\
  0 &  0 &  0 & -1 \\
  0 &  0 &  0 &  0 
\end{array}
\right], \label{sp2a2} \\
&&
H_{\alpha_4} ~=~
\frac{1}{2}\left[
\begin{array}{cccc}
  1 &    &    &    \\
    &  1 &    &    \\
    &    & -1 &    \\
    &    &    & -1 
\end{array}
\right], \quad 
e_{+\alpha_4} ~=~
\frac{1}{\sqrt{2}}\left[
\begin{array}{cccc}
  0 &  0 &  1 &  0 \\
  0 &  0 &  0 &  1 \\
  0 &  0 &  0 &  0 \\
  0 &  0 &  0 &  0 
\end{array}
\right]. \label{sp2a4}
\eea
The lowering operators $e_{-\alpha_i}$'s are given by hermitian 
conjugations of $e_{+\alpha_i}$'s above. 

\subsection{\large \bf G$_2$}

By using the 7$\times$7 $\rSO(7)$ angular momentum matrices 
$(M_{ab})_{cd} \equiv -i(\delta_{ac}\delta_{bd}-\delta_{ad}\delta_{bc})$,
fourteen generators of $G_2$ as a subset of $M_{ab}$'s are given by \cite{GG}
\bea
&& F_1 = M_{24}-M_{51},\quad 
   M_1 =-\frac{1}{\sqrt{3}}(M_{24}+M_{51}-2M_{73}),\nonumber\\
&& F_2 = M_{12}-M_{54},\quad 
   M_2 =\frac{1}{\sqrt{3}}(M_{54}+M_{12}-2M_{67}),\nonumber\\
&& F_3 = M_{14}-M_{25},\quad 
   M_3 =-\frac{1}{\sqrt{3}}(M_{14}+M_{25}-2M_{36}),\nonumber\\
&& F_4 = M_{16}-M_{43},\quad 
   M_4 =-\frac{1}{\sqrt{3}}(M_{16}+M_{43}-2M_{72}),\nonumber\\
&& F_5 = M_{46}-M_{31},\quad 
   M_5 =-\frac{1}{\sqrt{3}}(M_{46}+M_{31}-2M_{57}),\nonumber\\
&& F_6 = M_{35}-M_{62},\quad 
   M_6 =-\frac{1}{\sqrt{3}}(M_{35}+M_{62}-2M_{71}),\nonumber\\
&& F_7 = M_{23}-M_{65},\quad 
   M_7 =\frac{1}{\sqrt{3}}(M_{65}+M_{23}-2M_{47}),
\eea
where $F_1,\dots,F_7$ and $F_8\equiv-M_3$ consist of eight generators of 
SU(3). In fact, diagonalizing those matrices by using a unitary matrix
\be
U_2 =\frac{1}{\sqrt{2}}\left[
\begin{array}{rr|c}
 iI &  I &  \\
-iI &  I &  \\ \hline
    &    &  \sqrt{2}  \\ 
\end{array}
\right],
\ee
with the 3$\times$3 identity matrix $I$, $F_1,\dots,F_8$ turn to
\be
U_2 F_a U_2^\dagger =\left[
\begin{array}{ccc}
\lambda_a &              &   \\ 
          & -\lambda_a^T &   \\ 
          &              & 0
\end{array}
\right],
\ee
where $\lambda_a$'s denote Gell-Mann matrices. 
One can see that ${\bf 7}$ of $G_2$ goes to ${\bf 3 +\bar{3}+1}$ in SU(3). 
Thus we define fourteen generators $T_a$ of $G_2$ by
\be
T_a \equiv \left\{
\begin{array}{l}
\dfrac{1}{\sqrt{2}}\,U_2 F_a U_2^\dagger \quad\mbox{for}\quad a=1,\dots,8, \\
\dfrac{1}{\sqrt{2}}\,U_2 M_{a-8}\,U_2^\dagger \quad\mbox{for}\quad a=9, 10, \\
\dfrac{1}{\sqrt{2}}\,U_2 M_{a-7}\,U_2^\dagger \quad\mbox{for}\quad 
a=11,\dots,14.
\end{array}
\right.
\ee
The raising and lowering operators for simple roots $\alpha_1$ and 
$\alpha_2$ are given by
\be
E_{\pm \alpha_1} \equiv \frac{1}{\sqrt{2}}\left(T_6 \mp i T_7\right),\quad
E_{\pm \alpha_2} \equiv \frac{1}{\sqrt{2}}\left(T_9 \mp i T_{10}\right).
\ee
By using them, the raising and lowering operators for non-simple short 
roots are given by
\be
E_{+\alpha_4} \equiv [E_{+\alpha_1},E_{+\alpha_2}],\quad
E_{+\alpha_6} \equiv \frac{\sqrt{3}}{2}\,[E_{+\alpha_4},E_{+\alpha_2}],
\ee
and their hermitian conjugations. By using the above operators, 
the raising and lowering operators for non-simple long roots are 
determined as
\be
E_{+\alpha_3} \equiv [E_{+\alpha_4},E_{+\alpha_6}],\quad
E_{+\alpha_5} \equiv [E_{+\alpha_6},E_{+\alpha_2}],
\ee
and their hermitian conjugations.


\end{document}